\documentclass[]{aastex631}

\shorttitle{Flux Rope Reconnection}
\shortauthors{Chen et al.}
\graphicspath{{./}}
\begin{document}

%\title{Witnessing Magnetic Reconnection in Superpenumbral Fibrils Around a Sunspot}

\title{Witnessing Magnetic Reconnection in Tangled Superpenumbral Fibrils Around a Sunspot}

\correspondingauthor{Hechao Chen, Hui Tian}
\email{hechao.chen@ynu.edu.cn, huitian@pku.edu.cn}

\author[0000-0001-7866-4358]{Hechao Chen}
\affiliation{School of Physics and Astronomy, Yunnan University, Kunming, 650500,  China}
\affiliation{School of Earth and Space Sciences, Peking University, Beijing, 100871, China}
\affiliation{Yunnan Key Laboratory of the Solar Physics and Space Science, Kunming, 650216, China}

\author[0000-0002-1369-1758]{Hui Tian}
\affiliation{School of Earth and Space Sciences, Peking University, Beijing, 100871, China}

\author{Eric R. Priest}
\affiliation{School of Mathematics and Statistics, University of St Andrews, St Andrews, KY16 9SS, UK}

\author{Christopher B. Prior}
\affiliation{Department of Mathematical Sciences, Durham University, Durham DH1 3LE, UK}

\author[0000-0002-7153-4304]{Chun Xia}
\affiliation{School of Earth and Space Sciences, Peking University, Beijing, 100871, China}
\affiliation{Yunnan Key Laboratory of the Solar Physics and Space Science, Kunming, 650216, China}

\author{Anthony R. Yeates}
\affiliation{Department of Mathematical Sciences, Durham University, Durham DH1 3LE, UK}

\author[0000-0003-2891-6267]{Xiaoli Yan}
\affiliation{Yunnan Observatories, Chinese Academy of Sciences, Kunming 650011, China}
\affiliation{Yunnan Key Laboratory of the Solar Physics and Space Science, Kunming, 650216, China}

\author[0000-0001-9491-699X]{Yadan Duan}
\affiliation{Yunnan Observatories, Chinese Academy of Sciences, Kunming 650011, China}
\affiliation{Yunnan Key Laboratory of the Solar Physics and Space Science, Kunming, 650216, China}

\author[0000-0003-4804-5673]{Zhenyong Hou}
\affiliation{School of Earth and Space Sciences, Peking University, Beijing, 100871, China}

\author[0000-0002-2358-5377]{Zhenghua Huang}
\affiliation{Institute of Science and Technology for Deep Space Exploration, Suzhou Campus, Nanjing University, Suzhou 215163, China}

\author{Oliver E. K. Rice}
\affiliation{Department of Mathematical Sciences, Durham University, Durham DH1 3LE, UK}

%% Note that the \and command from previous versions of AASTeX is now
%% depreciated in this version as it is no longer necessary. AASTeX 
%% automatically takes care of all commas and "and"s between authors names.

%% AASTeX 6.31 has the new \collaboration and \nocollaboration commands to
%% provide the collaboration status of a group of authors. These commands 
%% can be used either before or after the list of corresponding authors. The
%% argument for \collaboration is the collaboration identifier. Authors are
%% encouraged to surround collaboration identifiers with ()s. The 
%% \nocollaboration command takes no argument and exists to indicate that
%% the nearby authors are not part of surrounding collaborations.

%% Mark off the abstract in the ``abstract'' environment. 
\begin{abstract}

Three-dimensional magnetic reconnection is a fundamental plasma process crucial for heating the solar corona and generating the solar wind, but resolving and characterizing it on the Sun remains challenging. Using high-quality data from the Chinese New Vacuum Solar Telescope, the Solar Dynamics Observatory, and the Interface Region Imaging Spectrograph, this  {work} presents highly suggestive direct imaging evidence of  magnetic reconnection during the untangling of braided magnetic structures above a sunspot. These magnetic structures, visible as bright superpenumbral threads in extreme ultraviolet passbands, initially bridge opposite-polarity magnetic fluxes and then gradually tangle in their middle section.  Magnetic extrapolation reveals the fibrils to form a small flux rope that is twisted and braided,  possibly created by persistent and complex photospheric motions. During untangling, repetitive reconnection events occur inside the flux rope, accompanied by transient plasma heating, bidirectional outflowing blobs,  {and signatures of nanojets}. Emission analysis reveals that the outflowing blobs are multi-thermal structures with temperatures well below 1 MK, undergoing rapid cooling and leaving emission imprints in H$\alpha$ images. The measured reconnection angles indicate that {16}$\%$-{22}$\%$  of the magnetic field along each thread is anti-parallel, with the remaining field acting as a guide field. The estimated energy released during these reconnection events is comparable to nanoflares, which can be powered by up to $6\%$ of the magnetic energy stored in the anti-parallel field. This work presents a textbook example of magnetic flux rope reconnection in the solar atmosphere,  providing new insights into fine-scale energy release processes within sunspot superpenumbral fibrils.

\end{abstract}

%% Keywords should appear after the \end{abstract} command. 
%% The AAS Journals now uses Unified Astronomy Thesaurus concepts:
%% https://astrothesaurus.org
%% You will be asked to selected these concepts during the submission process
%% but this old "keyword" functionality is maintained in case authors want
%% to include these concepts in their preprints.

%% From the front matter, we move on to the body of the paper.
%% Sections are demarcated by \section and \subsection, respectively.
%% Observe the use of the LaTeX \label
%% command after the \subsection to give a symbolic KEY to the
%% subsection for cross-referencing in a \ref command.
%% You can use LaTeX's \ref and \label commands to keep track of
%% cross-references to sections, equations, tables, and figures.
%% That way, if you change the order of any elements, LaTeX will
%% automatically renumber them.
%%
%% We recommend that authors also use the natbib \citep
%% and \citet commands to identify citations.  The citations are
%% tied to the reference list via symbolic KEYs. The KEY corresponds
%% to the KEY in the \bibitem in the reference list below. 

\section{Introduction} \label{sec:intro}  
Stellar magnetic fields are widely believed to play a central role in energizing and maintaining the million-Kelvin hot coronae for cool stars, including our Sun and M dwarfs. However, decoding the detailed mechanism of converting magnetic energy into thermal energy still remains one of the outstanding open issues in astrophysics. On the Sun, the dissipation of magnetohydrodynamic waves \citep{2011ApJ...736....3V,2013SSRv..175....1M,2020SSRv..216..140V} and the release of magnetic stresses through magnetic reconnection \citep{1972ApJ...174..499P,1983ApJ...264..642P,1988ApJ...330..474P} are thought to be leading physical processes to heat the corona \citep{2006SoPh..234...41K,2014LRSP...11....4R}.  As one of prime candidates for coronal heating,  Parker’s nanoflare theory posits that coronal magnetic fields rooted in the photosphere are subject to random and turbulent footpoint motions \citep{1988ApJ...330..474P} . This can lead to the braiding {or shearing} of magnetic field lines, injecting Poynting flux into the corona and steadily increasing the amount of free magnetic energy stored in the coronal magnetic field. As magnetic field gradients and currents accumulate within the {twisted or} braided field lines, tiny reconnection processes between the anti-parallel components of the magnetic field would release the stored magnetic energy; this energy can be dissipated by collisionless or resistive currents, and involve particle acceleration, plasma outflow jets, MHD wave excitation, and plasma viscosity, thereby thermalizing the local corona \citep{2000mare.book.....P,2015ApJ...811..106H,2022LRSP...19....1P}.

The role of footpoint motions and their associated potential for coronal heating {has} been extensively explored in various theoretical and numerical heating models \citep{1994ApJ...422..381C,2001ApJ...553..440K,2004ApJ...617L..85P,2002ApJ...576..533P,2008ApJ...677.1348R,2015RSPTA.37340260C,2015ApJ...811..106H}. These leading models consistently predict that nanoflares release energy on the order of 10$^{24}$ erg, over very small temporal and spatial scales. 
Direct imaging validation of the weak and transient coronal heating signals associated with nanoflares would be invaluable for addressing the coronal heating problem, but it has been challenging with early-era solar remote sensing instruments. %This difficulty is further compounded by observational factors, such as projection and obscuration effects, which can hinder the unambiguous detection of these subtle signals in many candidate nanoflare events. %Some geophysical observations and theoretical studies suggest that the reconnection rate for antiparallel reconnection appears to be higher than that for component reconnection \cite{2004PhPl...11.5387H,2007JGRA..11210206M}. 
Three-dimensional reconnection can occur at a null point, a separator, a quasi-separator or by braiding {or twisting} \citep{2022LRSP...19....1P,2023AdSpR..71.1856P}. In the {braiding \citep{1972ApJ...174..499P,2020LRSP...17....5P} or tectonics \citep{2002ApJ...576..533P} model}, reconnection occurs mainly at nearly parallel fields at small shear angles with a strong guide field. The precise impact of a guide field remains controversial \citep{2004JGRA..109.1220P,2006GeoRL..3313105D,2010ApJ...725..319Q,2015ApJ...799...79N,2016ApJ...818...20H,2022NatCo..13.6426B}. This fuels ongoing debate about the detailed characteristics of braiding or tectonics and their contribution to coronal heating, which in turn stresses the need for more constraints from observations of the solar atmosphere.

To date, there are only a few sporadic and fragmented solar imaging clues about the occurrence of tiny flux rope reconnection in the solar atmosphere {associated with coronal heating}. In the past, low-intensity bursts within the nanoflare energy range, i.e., transition region (TR) subarcsecond bright dots above sunspots \citep{2014ApJ...790L..29T,2016ApJ...829..103D,2017ApJ...835L..19S}, penumbral chromospheric microjets \citep{2007Sci...318.1594K,2019A&A...626A..62R}, impulsive moss brightenings at the footpoints of hot dynamic loops \citep[e.g.,][]{2013ApJ...770L...1T,2014Sci...346B.315T}, as well as recently reported campfires in the quiet corona \citep[e.g.,][]{2021A&A...656L...4B,2021A&A...656L...7C}, have been hypothesized to be byproducts of possible such reconnection. Moreover, the morphological evidence of magnetic braids, when associated with heating signals, is often considered a compelling indicator of reconnection in candidate nanoflare events \citep{2013Natur.493..501C,2022A&A...667A.166C,2023A&A...679A...9B}. The high spatial resolution and high cadence observations from the Extreme Ultraviolet Imager on board Solar Orbiter recently provided new evidence for the untangling of small-scale coronal braids in the core of active regions, where the untangling of magnetic fields was found to be associated with gentle or impulsive heating of the gas in related coronal loops \citep{2022A&A...667A.166C}. Moreover, the Interface Region Imaging Spectrometer (IRIS)'s TR imaging observations revealed very fast and bursty mini-jets occurring roughly perpendicular to the magnetic field in activated prominences \citep{2020ApJ...899...19C,2021NatAs...5...54A}. \citet{2021NatAs...5...54A} termed these tiny jet phenomena ``nanojets" and interpreted them as strong evidence for reconnection-based nanoflares \citep[also see][]{2022ApJ...938..122P,2022ApJ...934..190S,2025ApJ...985L..12G}. However, due to the weak and transient nature of the observed signals, the detailed physical process of reconnection and braiding {is} difficult to resolve in most of these aforementioned observations.

In this  {work}, utilizing high-quality imaging observations from various solar telescopes, we present direct imaging evidence of a series of repetitive reconnection events occurring during the untangling of  magnetic fibrils in the solar chromospheric and transient-region (TR). For the first time, our research resolves the relatively complete tangling and untangling of magnetic-field structures around a sunspot, and characterizes the thermal, magnetic, and dynamic nature of such reconnection by capturing their distinct heating and plasma flow signatures. This study offers valuable new insights into reconnection involved in the  untangling of magnetic flux ropes on the Sun.

\section{Observations} \label{sec:obs}

The repetitive {reconnection} events were observed in the solar atmosphere above an $\alpha$-type sunspot in active region (AR) NOAA 12833 on 2021 June 21. High-resolution and multi-wavelength imaging observations from the New Vacuum Solar Telescope (NVST) \citep{2014RAA....14..705L}, Interface Region Imaging Spectrograph (IRIS) \citep{2014SoPh..289.2733D}, and Solar Dynamics Observatory (SDO) \citep{2012SoPh..275....3P} were used to jointly observe these weak reconnection signals, which were characterized as bursty and transient brightenings in extreme-ultraviolet (EUV) and ultraviolet (UV) images occurring intermittently from 07:00 UT to 18:00 UT.  We primarily investigated {flux rope} reconnection events during 09:20$-$10:10 UT and 17:50$-$18:00 UT, which were well-observed by NVST and SDO, or IRIS and SDO, respectively. 

The NVST tracked AR 12833 from 09:22:11 UT to 10:19:45 UT in the H$\alpha$ line core (6563 Å) and TiO, with high temporal resolution (12 s) and  {a pixel size of about} 0.165$^{\prime\prime}$ for H$\alpha$ and 0.052$^{\prime\prime}$ for TiO \citep{2014RAA....14..705L,2020ScChE..63.1656Y}. These NVST images were calibrated through dark current and flat field corrections and further reconstructed using a speckle masking method  \citep{2016NewA...49....8X}. {IRIS} tracked AR 12833 from 17:00:31 UT to 23:05:04 UT with  a large coarse 8-step raster (119$^{\prime\prime}$ along the slit, eight raster steps with a 2$^{\prime\prime}$ step size).  {IRIS also provided high-resolution slit-jaw images (SJIs) with a field-of-view (FOV) of 120$^{\prime\prime}\times$119$^{\prime\prime}$ in the 1400 Å, 1330 Å, and 2794 Å passbands. These IRIS data had a time cadence of 20 s, an exposure time of 4 s, and a pixel size of approximately 0.33$^{\prime\prime}$}. Calibrated level-2 IRIS data, which has undergone dark-current subtraction, flat-fielding, and geometrical correction, were utilized \citep{2014SoPh..289.2733D}. 

The Atmospheric Imaging Assembly (AIA) \citep{2012SoPh..275...17L}  on board the SDO observes the full solar disk in seven EUV passbands, simultaneously sampling plasma at different temperatures ranging from 0.6 to 20 MK. Six AIA Fe-line filters capture EUV emission Fe lines formed at 94 \AA\ (Fe \textsc{xviii}), 131 \AA\ (Fe \textsc{viii}, Fe \textsc{xx}, Fe \textsc{xxiii}), 171 \AA\ (Fe \textsc{ix}), 193 \AA\ (Fe \textsc{xii}, Fe \textsc{xxiv}), 211 \AA\ (Fe \textsc{xiv}), and 335 \AA\ (Fe \textsc{xvi}). The AIA 304 filter reveals structures in the cooler chromosphere and transition region around log$_{10}$ T (K) of 4.7, sampled by He \textsc{ii} in AR  \citep{2010A&A...521A..21O}. 
The spatial pixel size is approximately 0.6$^{\prime\prime}$ with a time cadence of 12 seconds. 
The pre-calibrated Active Region Patches (SHARPs) product \citep{2014SoPh..289.3549B}, provided by the Helioseismic and Magnetic Imager (HMI) \citep{2014SoPh..289.3483H} on board SDO, were utilized to study the magnetic topology of AR 12833. These data have a spatial pixel size of 0.$^{\prime\prime}$5.
AIA images were calibrated using the standard aia$\_$prep.pro routine in the SolarSoftWare IDL package. All AIA, IRIS, and NVST data were accurately co-aligned by matching corresponding sunspot structures in the AIA 1700 Å, SJI 1330 Å, and TiO images. To compensate for the solar rotation, all the images taken at different times are aligned to an appropriate reference time.

\section{Results and Discussion} \label{sec:result}

\subsection{Magnetic {Flux Rope} Reconnection}
The repetitive reconnection events took place along a bundle of  superpenumbral fibrils extending outward across the moat of the $\alpha$-type sunspot to a nearby network region (Figure \ref{fig1}a). These superpenumbral fibrils, with an apparent length of 28$-$36 Mm, consist of both bright and dark fine threadlike structures. They are well resolved as bright threads in AIA 171 \AA~images and/or dark threads in NVST H$\alpha$ images, respectively (Figures \ref{fig1}b and  \ref{fig1}c). Hereafter, we refer to these plasma threadlike structures as bright threads for clarity. By visually identifying their footpoints, we can see that these bright threads root most of their north footpoints in the positive-polarity sunspot, with some anchoring to a positive-polarity network-field region; all south footpoints of these bright threads were anchored in a negative-polarity network-field region (see the HMI vector magnetograms and NVST TiO images in Figure \ref{fig1}).  This suggests that the magnetic field of these bright threads were initially distributed along an almost parallel orientation. 

The reconnection repetitively took place among these nearly parallel bright threads as they gradually {twisted and} intertwined in the middle section. A prime example of this process is evident in the AIA 171 \AA~ images of Figure \ref{fig1}a. Initially, at 09:28:21 UT, several bright threads displayed a weak twisted and braided morphology at their intersection point, with a small crossing angle of approximately 18 to 25 degrees on the plane of the sky. By 09:46:45 UT, this morphology become more and more pronounced, with at least three bright threads spatially resolved and involved.  Such a morphology of coronal loops has been regarded as a slender magnetic flux rope in previous studies \citep{2013Natur.493..501C,2014ApJ...780..102T,2018ApJ...854...80H}. Approximately one minute later, enhanced emissions emerged from the braided knot of these bright threads, forming a chain-like brightening region at about 09:48:09 UT. From this region, two distinct bright blobs rapidly developed and streamed outward along the two ends of the bright threads within 40 seconds. This suggests the presence of magnetic reconnection  and its associated dynamics.  Furthermore, the apparent blob-like morphology of these bidirectional plasma flows is similar to plasma blobs extensively reported in various solar activities, which may be signatures of plasmoids generated by magnetic reconnection \citep[e.g.,][]{2016NatPh..12..847L,2017ApJ...851L...6R,2019ApJ...879...74C,2022NatCo..13..640Y,2023NatCo..14.2107C,2024A&A...687A.190H,2023A&A...673A..11R,2025ApJ...979..195D,2025A&A...696A...3L}. 

The NVST also captured this episode of reconnection. As Figure \ref{fig1}c illustrates, the running-difference H$\alpha$ image taken at 09:41:01 UT also reveals a pronounced twisted morphology of bright threads, highlighted by a yellow arrow. Notably, these reconnection episodes occur near the midpoints of twisted bright threads, without obvious transfer of twist or mass to the ambient field. Such features strongly imply that the reconnection is primarily initiated within the flux rope itself, as opposed to occurring in the interlayer between the flux rope and the less-sheared ambient field. The EUV bidirectional blobs observed in the 171  \AA~images are manifested as weak and transient bright front features superimposed on the dark superpenumbral fibrils in the H$\alpha$ images and the accompanying animation. These unique H$\alpha$ bright front features exhibit continuous and rapid outward streaming along both sides of the {twisted} structure. 

Morever, two episodes of similar repetitive reconnection are captured in AIA 304 \AA~images, as presented in Figures \ref{fig2} (a1)-(a5) and (b1)-(b5). Initially, bidirectional blob outflows first emerge from a very tiny bright dot, as seen in Figures \ref{fig2} (a1) and (b1). This bright dot appears at a similar scale to the reported IRIS TR subarcsecond bright dots above sunspots \citep{2014ApJ...790L..29T}. These newly-born blobs then grow and propagate along bright threads, tracing out their north footpoints at the sunspot region and clear bifurcated footpoints at their southern end. As seen seen in Figures \ref{fig2} (b4) and (b5), such a bifurcated footpoint feature is consistent with the flux rope scenario for the magnetic field. The observed emission enhancements in the AIA 304, 171, and H$\alpha$ passbands are thus most likely driven by magnetic reconnection occurring at small angles between the twisted or braided bright threads inside the flux rope. The apparent angles between the crossing threads were approximately 25 degrees.  {Furthermore, a closer look at the reconnection site at the midpoints of the twisted bright threads revealed signatures of ongoing nanojets (see Appendix A). These weak nanojets, which were only discernible in 304 \AA~images, provide strong evidence for the occurrence of flux rope reconnection. Two examples are shown in Figure \ref{figA0} and its associated animation. They are characterized by short-lived ($\sim$ 60-70 s) and rapid-moving ($\sim$ 40-60 km s$^{-1}$) jet-like plasma ejecta. Similar to  the recently reported mini-jets \citep{2020ApJ...899...19C,2025arXiv250904741T} or nanojets \citep{2021NatAs...5...54A,2022ApJ...938..122P,2022ApJ...934..190S,2025ApJ...985L..12G}, as well as some cases of  plasma blob ejection \citep{2025ApJ...985...17Z}, they have a clear velocity component perpendicular to the reconnecting bright threads.}

\subsection{Outflowing Blobs: Dynamics and Emissions}
 {We mainly focus on the bidirectional outflowing blobs originating from the magnetic reconnection site.} To better characterize these outflowing blobs, time-distance diagrams were constructed using AIA images at 171 and 304  \AA~along the slice AB, as shown in the third panel of Figure {\ref{fig1}b}. From Figures \ref{fig3}c and \ref{fig3}d, it is clear that these outflowing blobs are repetitive phenomena, manifesting as pairs of stripes that rapidly outflow from their {reconnection} site. Note that these outflows primarily start in the middle section of the slice AB, spatially corresponding to the middle section of the {twisted} bright threads where they intertwine. These outflows are even more prominent in the 304 \AA~ passband  than in 171 \AA. As shown in Figures \ref{fig2}(e)-(g), our statistical analysis of all the identified 304 \AA~outflowing blobs reveals that: (1) their projection propagation velocities range from  20 to 230 km s$^{-1}$, with an average value ($\overline{v}$) of 95 km s$^{-1}$;  (2) their propagation duration ranges from 50 to 360 s, with an average value of 140 s; (3) their propagation distance spans from $\sim$ 4 to 20 Mm, with a positive correlation with their outflowing velocity.  {Moreover, most of the outflowing blobs flow along the untangling bright threads. This behavior differs significantly from  {the nanojets we previously observed} and also contrasts with some cases of plasma blob ejection \citep{2025ApJ...985...17Z} that are ejected perpendicular to the reconnection field line. The observed outflowing blobs in this study are likely to be reconnection-induced plasma flows driven by magnetic tension forces and pressure gradients along magnetic field lines \citep{2013Natur.493..501C}, since they are located far from the reconnection diffusion region. The diffusion region of flux rope reconnection is believed to occur within very narrow regions of strong electric current \citep{2000mare.book.....P,2022LRSP...19....1P}, which are not  spatially resolved by current observational instruments.

Using a threshold-based identification method, we identified a total of 35 independent outflowing blobs at their maximum intensity using EUV 304 and 171 \AA~images taken between 09:20 and 10:10 UT, respectively.  This technique was accomplished by identifying image pixels where the blob intensity exceeded 3$\sigma$ above the median intensity along the chain-like bright region. As illustrated in Figure \ref{fig3} (a1), this threshold ensures that all the brightest pixels encompassing the bright blob region can be distinguished and separated from the noisy background.
During this period of time, we obtained their  apparent area, length ($h$), and width ($d$). The spatial scale of these outflowing bright blobs is similar to the mini-jets \citep{2020ApJ...899...19C}.  The identified bright blobs have an average apparent area in the 304 Å passband that is larger than in the 171 Å passband, with values of 4.61 Mm$^2$ and 2.50 Mm$^2$, respectively (see Figure \ref{fig3}d). Their average apparent length-to-width ratio is 3.48/1.32 = 2.63 in the 304 Å passband and 2.09/1.19 = 1.75 in the 171 Å passband (see Figure \ref{fig3}e). This result suggests that the outflowing bright blobs are elongated, multi-thermal structures with a compact hotter core enveloped by an elongated cooler shell. This spatially-dependent thermal nature of these bright blobs is reminiscent of the multi-thermal internal structures, a hot core inside a cool outer layer structure, as observed in coronal rain \citep[e.g.,][]{2015ApJ...806...81A,2022A&A...659A.107C}. This striking similarity suggests that the initial hot plasma within these outflowing blobs undergoes a cooling process after formation, thus leaving unique emission imprints over {the} NVST H$\alpha$  superpenumbral fibrils.

These outflowing bright blobs also appear in other EUV passbands. As shown in Figure \ref{fig3}, the blobs are relatively prominent in the 131, 193, 211, and 335 Å images, but least visible in the 94 Å passband possibly due to its lower signal-to-noise ratio. By focusing on the chain-like brightening region where the outflowing blobs emerge (outlined by the dotted lines in Figure \ref{fig3} (a3)), the normalized flux intensities of six sensitive AIA passbands were examined. As illustrated in Figure \ref{fig3}b, six distinct flux intensity enhancements recurrently appear as sudden peaks between 09:40 and 10:30 UT, with a time interval of approximately 6 to 8 minutes. These intensity enhancements do not exhibit a significant time delay across different AIA passbands, suggesting that their emissions may originate from a similar temperature range. 
Given the obvious response around 10$^{5.5}$ K across all these AIA passbands, it is highly probable that these bright blobs represent cool features at TR temperatures, resembling the bidirectional propagating blob-like features recently reported in Solar Orbiter/EUI observations \citep[e.g.,][]{2024A&A...692A.119C}. We conducted EM-loci measurements of these bright blobs using 131, 171, 93, 211, and 335 Å channels, due to their relatively higher signal-to-noise ratio (see Figure \ref{fig4}c). In this technique, the temperature response function of all selected channels was calculated; EM-loci curves for each selected channel were then generated by dividing the background-subtracted average intensity from all pixel positions of the identified blobs at their peak time by the response functions \citep{2002A&A...385..968D,2013ApJ...771...21W,2014ApJ...790L..29T}.
Figure \ref{fig4}c displays the EM-loci curves of one typical bright blob. Assuming the initial plasma within the newborn outflowing blobs is isothermal, the upper limit plasma temperature and EM of the blob is measured at $10^{5.47 \pm 0.13}$ K and (8.15 $\pm$ 3.35) $\times$ 10$^{27}$ cm$^{-5}$, respectively. 
Using the same EM-loci analysis, the average upper limit of plasma temperature ($\overline{T}$) for all identified bright blobs was determined to be $10^{5.43 \pm 0.03}$ K, with an average emission measure ($\overline{EM}$) of 5.27 ± 1.75 × 10$^{27}$ cm$^{-5}$ (see Figures \ref{fig3}(f) and (g)).

\subsection{Energy Budget}
 If the line-of-sight integral thickness of typical bright blobs equals their average width ($\overline{d} \sim 1.32 $ Mm in 171 \AA~image), their electron density can be calculated as $n_e = \sqrt{(\overline{EM})/(\overline{d})} \approx 6.32 \times 10^{9}$ cm$^{-3}$ . We then calculated the average thermal and kinetic energy densities ($\overline{E_t} = 3n_ek_B\overline{T}$ and $\overline{E_k} = \frac{1}{2}n_em_p\overline{v}^2$ ) for typical outflowing blobs, finding values of 0.65 and 0.48 erg cm$^{-3}$, respectively.  If we assume that thermal and kinetic energy dominate over other forms of energy, their sum roughly {implies} a dissipated magnetic energy ($\overline{E_m}$) of approximately 1.13 erg cm$^{-3}$. Based on the magnetic energy density formula $E_m = (\Delta B)^2/{8\pi}$,where $\Delta B$ is the oppositely directed (i.e., antiparallel) field component involved in the reconnection, it is speculated that approximately 5.3 G of magnetic field strength ($\Delta B$) along the braided bright threads is converted into thermal and kinetic energy on average during the reconnection process. Moreover, considering the typical bright blob as a cylinder of plasma with a volume of $\pi (\overline{d}/2)^2\cdot \overline{h}$, {an upper limit on the} total dissipated magnetic energy {can thus} be computed as $5.40 \times 10^{24}$ erg on average, comparable to the typical nanoflare energy level of $\sim$ $10^{24}$ erg \citep{1988ApJ...330..474P}. This is also close to the reported energy estimates of other weak nanoflare  candidates, i.e.,  about $10^{22}- 10^{23}$ erg for TR subarcsecond bright dots above sunspots \citep{2014ApJ...790L..29T} and $\sim 10^{24}$ erg for nanojets in activated solar prominences \citep{2020ApJ...899...19C,2021NatAs...5...54A}. 

\subsection{Magnetic field Modeling and Topology Analysis}
Direct and accurate measurements of the magnetic field in the Sun’s active-region corona remain challenging, despite recent progress in understanding the magnetic field of the global corona achieved through coronal seismology \citep{2020ScChE..63.2357Y,2020Sci...369..694Y,2024Sci...386...76Y}.
Therefore, we constructed a non-linear force-free field (NLFFF) model to approximate the three-dimensional (3D) coronal magnetic configuration of AR 12833. This was achieved using the NLFFF package available in SolarSoft (SSW) and the “weighted optimization" technique \citep{2000ApJ...540.1150W,2004SoPh..219...87W}. A pre-calibrated SHARPs vector magnetogram taken at 06:12 UT (see Figures \ref{fig4} and \ref{figA1}) serves as the lower boundary for the NLFFF model, covering a field of view with solar xrange = [398$^{\prime\prime}$, 693$^{\prime\prime}$] $\times$ yrange = [35$^{\prime\prime}$, 410$^{\prime\prime}$]. Prior to extrapolation, this input bottom boundary was ``preprocessed" to best suit the force-free condition \citep{2006SoPh..233..215W}.
As depicted in Figures \ref{fig4}(a) and (b), a set of field lines indicated in green, traced and selected from their negative-polarity footpoint region, is present at the location where the bright threads were observed. As shown in Figures \ref{fig4}(c1) and (c2), this bundle of field lines displays obvious entanglement in its middle section, which aligns with the EUV imaging findings in Figure \ref{fig1}(b). 
We further demonstrate that this bundle of field lines has significantly complex topology by calculating the mean twisting ${\cal T}_w(\gamma_i,\gamma_j)$ of all pairs of fieldlines $(\gamma_{i}, \gamma_{j})$. This was calculated by evaluating the net-winding (linking) $L(\gamma_i, \gamma_j)$ of each pair of curves and the writhe (self-entanglement) $W(\gamma_i)$ of each curve $\gamma_i$ (see \citet{2006JPhA...39.8321B} for details).
As shown in \citep{2006JPhA...39.8321B}, ${\cal T}_w(\gamma_i,\gamma_j) = L(\gamma_i,\gamma_j) -W(\gamma_i)$.
%$\gamma_j$ around $\gamma_i$ is the difference $ L(\gamma_i, \gamma_j)-W(\gamma_i)$.  
This is a precise means of calculating the twisting of pairs of curves in the extrapolated flux rope. Applying this calculation to all pairs of  curves in the indicated set in Figure \ref{fig4}(b) yielded a (dimensionless) mean twisting ${\cal T}_w$ of 0.641 (to 3.s.f), with a standard deviation of 0.531 (to 3.s.f). A twist of $0.5$ is (net) one half of a right-handed turn of the pair of curves. So, there is a significant net positive twist in the field, but the magnitude of the standard deviation also indicates significant variation in the pairwise entanglement, a sign of more complex braided structure (both positive and negative entanglement) on top of this net twist. 

To investigate the whole field’s topological structure more thoroughly we calculate the distributions of the fieldline helicity (using the winding gauge\citep{2014ApJ...787..100P,2025RSPSA.48140152X}), the mean fieldline twisting, and the fieldline integrated current helicity. 
All quantities were calculated from the extrapolated magnetic fields, and then mapped to their corresponding photospheric footpoints, as presented in Figures  \ref{fig4}(d)-(f). These topological metrics are very helpful for characterizing the complexity of solar active region magnetic fields  \citep[e.g.,][]{2012ApJ...752L...9J,2021NatCo..12.6621M,2022ApJ...927..156R,2025ApJ...980..102W}. 
What is interesting is that the set of  green field lines shown in figures \ref{fig4}(a) and (b)  do not have the same topology as the main  sheared arcade, which joins the main positive pole to the southern negative flux patch. 
This can be observed in panel  \ref{fig4}(d), which shows that the field line helicity (the flux-weighted mean entanglement of each field line) is dominantly negative in these flux regions. 
 To explore this further we calculate in Figure \ref{fig4}(e) the following proxy of the mean twisting $T(\gamma)$ of a fieldline $\gamma$ : 
\begin{equation} 
T(\gamma) = \int_{\gamma}\frac{\nabla \times {\bf B}\cdot {\bf B}}{{\bf B}\cdot{\bf B}} \mathrm{d}s .
\end{equation}
As shown in, \citet{2006JPhA...39.8321B}, \citet{2016ApJ...818..148L}, and \citet{2022GApFD.116..321P}, this quantity generally represents the mean twisting of the local field around the fieldline $\gamma$ which is integrated over. However, if there is significant local divergence of field lines (e.g. QSL’s or Null points) this also contributes to its value. In short $ T(\gamma) $ is a dimensionless measure of the local complexity of the field. 
We see in Figure \ref{fig4}(e) that the region from which the field lines are launched have significant complexity with values of both positive and negative sign, whilst it does not have such a large value in the high field line helicity regions seen in Figure \ref{fig4}(d). 
This indicates that the helicity is dominated by a weakly sheared arcade of negative chirality (the strong flux is what leads to a high helicity value). 
 By contrast the mixed sign complexity of $ T(\gamma) $ the region of interest (where the field lines are rooted) is commensurate with the pairwise topological calculations of the field line twisting ${\cal T}_w$ (a geometrically precise measure of twisting), which showed a significant net positive twist with significant additional mixed-sign entanglement. We interpret the difference in the two distributions as indicating a sheared arcade which has been substantially changed by reconnection and/or photospheric flows around its core footpoints (which, by contrast, retain the initial topology). 
 
Finally, in Figure \ref{fig4}(f)  we see the field lines shown in (a) and (b) have significant current helicity $H_c(\gamma)$ of a field line $\gamma$, given by the formula: 
\begin{equation} 
H_c(\gamma)= \int_{\gamma}(\nabla \times {\bf B}\cdot {\bf B}) \mathrm{d}s,
 \end{equation}
which is the flux weighted mean twisting of the field \citep{2012ApJ...752L...9J}.  It is proportional to the dot product of the current and the magnetic field. Again, this has significant values in the region of interest. So, we have identified a complex positively twisted and braided set of field lines, in a localised region, which facilitates the accumulation of electric current at a fine scale within the flux rope. The classical theory of nanoflare activity, induced by either braiding \citep{1972ApJ...174..499P,2020LRSP...17....5P} or tectonics \citep{2002ApJ...576..533P}, suggests that the electric current accumulation may favor the dissipation of magnetic energy via fine-scale three-dimensional reconnection \citep{2020LRSP...17....5P,2022LRSP...19....1P}.
These results suggest as a possible interpretation that a negative flux rope emerged to form the sunspot region, but during its development either reconnection and/or photospheric flows led to the development of a more complex mixed topology in the sub-region where our observations are located. We investigate this hypothesis by a further analysis of the observational data in the following section. 

In the NLFFF model, this flux rope  has an average length of 35 Mm, reaching up to 8.8 Mm in the Z-direction. As seen in Figures \ref{figA1} (a)) and \ref{fig4}(b), it is situated beneath less sheared background arcade fields originating from the sunspot. In accordance with the EUV and UV imaging findings shown in Figures \ref{fig1}(b) and \ref{fig5}(b), the extrapolated flux rope displays winding in its middle section, with its north footpoints rooted at the positive-polarity sunspot boundary (with an averaged magnetic field strength of 880 G) and its south footpoints extending towards the negative-polarity network field region  (with an averaged magnetic field strength of 270 G). Numerous crossing knots are seen along the intertwined bright threads (highlighted by arrows in Figures \ref{fig4} (c1) and (c2)). The magnetic field strength near crossing knots ranges from 137 to 180 G. Given that the small angles between the twisted bright threads span $18^\circ$ to $25^\circ$ on the plane of sky, the strengths of the anti-parallel magnetic field component and the guide field for each bright thread are roughly $21.{4}$ G to $39.0$ G and ${135.3}$ G to ${175.7}$ G, respectively. Based on our magnetic energy budget estimation, the magnetic strength dissipated along two braided threads is approximately 5.3 G, accounting for {$13.6\%-24.7\%$} of the strength of the anti-parallel magnetic field component for intertwined bright threads indicated in Figures \ref{fig1}(b) and \ref{fig4}(b), assuming they have identical magnetic field strengths. Therefore,  {$1.9\%-6.1\%$} of the magnetic energy stored in the anti-parallel component along these flux rope threads is sufficient to power the magnetic energy dissipation occurring in flux rope reconnection. 

\subsection{Recurrent Presence and Possible Origin of Flux Rope Topology}
The subsequent IRIS observations from 17:00:31 UT to 23:05:04 UT further confirmed the recurrent presence of intertwined magnetic structures in the same region. As depicted in Figure \ref{fig5}, the joint imaging observation from IRIS and AIA around 18:00 UT reveals another episode of reconnection occurring as  a similar crossing of bright threads happened. Consistent with the NLFFF model, a crossing knot, as marked by white arrows, appeared at the middle section of two intertwined bright threads. In UV 1400 \AA~images, this shearing-to-unshearing process rapidly took place within a few minutes, during which the bright threads transformed from an almost-parallel morphology into a crossing state, and then reverted back to an almost parallel morphology.  In 171 \AA~images, the flux rope of bright threads is poorly observed, instead appearing as a bright region that closely resembles the bright blobs mentioned earlier. During this magnetic tangling-to-untangling process, noticeable plasma heating quickly occurred over a time of approximately three minutes. The transient and subtle enhancement in emission indicates a rapid onset and cessation of the  reconnection process along two crossing bright threads. After the reconnection, they quickly reverted to a nearly parallel state within 60 seconds. Similarly, this might imply that the repetitive reconnection observed earlier likely took place at various braid knots along different intertwined bright threads.  

The observed crossing followed by uncrossing of magnetic-field structures, visible as bright threads in EUV 171 Å and 304 Å  images, has in principle two possible causes, namely twisting footpoint motions followed by either untwisting motions or reconnection. In our case, it is the accompanying plasma heating that makes the second explanation far more likely and so that is what we assume here. The slow and persistent photospheric motions at the footpoints of  magnetic flux tubes are thought to create magnetic flux rope or braids, depending on the complexity of the motions and provided that reconnection does not destroy them during the establishment of a braid. Near the flux rope's southern footpoint, we found persistent and  complex flow patterns that varied over shorter time periods (see Appendix B and Figure \ref{figA1}). This complexity of footpoint motions could effectively create a braided flux rope. 
Our observations of crossing structures with NVST is suggestive and our topological analysis suggests that the flux rope is twisted and braided. In this case new reconnections may be triggered along the crossing threads as electric current accumulates along the magnetic field lines  \citep{1988ApJ...330..474P}. This scenario is consistent with the classical nanoflare theory by braiding or tectonics and can elucidate the observed repetitive occurrence of  flux rope creation and subsequent reconnection of the crossing bright threads.

The energy flux input ($F$) resulting from the gradual shuffling and intermixing of magnetic-field footpoints by photospheric motions can be estimated using the equation \citep{1988ApJ...330..474P}: 
$F = \frac{B^2}{(4\pi L)}v_h^2t$, {in units of} ergs m$^{-2}$ s$^{-1}$. For an active region, the total radiative and conductive energy losses from the corona amount to an energy flux of approximately $10^7$ ergs cm$^{-2}$ s$^{-1}$ \citep{2006SoPh..234...41K}. With a typical magnetic field strength of $B = 10^2$ G, a typical horizontal velocity of $v_h = 1$ km s$^{-1}$, and an average bright length of $L = 30$ Mm for the braided bright threads in our observations, the energy flux can be estimated to reach the required level in a timescale of $t = 3725$ s, {or} approximately 1 hour. In our observations, the braided magnetic topologies recurrently appear in the same region at 06:12 UT, 09:40 UT, and 18:00 UT. The time intervals between each appearance seems sufficient for the necessary energy accumulation. 

\section{Summary} \label{sec:Sum}

Using high-quality data from the NVST, the SDO, and the IRIS, we present evidence for magnetic reconnection during the untangling of a flux rope consisting of superpenumbral fibrils around the sunspot in AR 12833. Through feature identification, dynamical analysis, and emission measurements, we characterize the thermal and dynamic aspects of this flux rope reconnection, focusing on its distinct heating and plasma flow signatures. Additionally, by combining magnetic field extrapolation with observations, we explore the flux rope's magnetic topology and its possible origin. The main findings are summarized as follow: 
\begin{itemize}
    \item Superpenumbral fibrils emanating from the sunspot, visible as bright threads in  H$\alpha$, EUV 171 Å and 304 Å  images, initially bridge two opposite-polarity magnetic fluxes in a nearly parallel configuration. Subsequently, they gradually intertwine at their midpoints at small angles and create a twisted and braided flux rope, possibly due to persistent and complex footpoint motions. Our magnetic extrapolation confirms the presence of a complex flux rope in the same region, exhibiting both twist and braiding.
    \item After its creation, the flux rope undergoes repetitive reconnection events at the midpoints of its crossing field lines, characterized by transient plasma heating, and rapid EUV outflowing blobs that occur along the field lines,  {as well as 304 \AA~nanojets perpendicular to the reconnecting field lines}. As a result, the twisted and braided bright threads of the flux rope transformed from a crossing state into an almost-parallel morphology within a few minutes. IRIS observations reveal the recurrent presence of this type of flux rope reconnection in similar flux rope structures. The energy level of these reconnection events is comparable to that of a typical nanoflare, approximately $10^{24}$ erg.    
        \item The crossing threads range from 18 to 25 degrees in most of the reconnection episodes, indicating that $15\%$ to $21.6\%$ of the magnetic field along each crossing thread is anti-parallel on the plane of the sky. Our estimate indicates that only $1.9\%-6.1\%$ of the magnetic energy stored in the anti-parallel component of the twisted magnetic field is sufficient to power such flux rope reconnection events. 
       \item  The striking outflow blobs generated by the flux rope reconnection are field-aligned, rapid plasma flows with an average velocity of 95 km s$^{-1}$. They are elongated, multi-thermal structures predominantly composed of warm TR plasma, with temperatures remaining below 1 MK. Each blob features a compact hotter core enveloped by an elongated cooler shell,  which undergoes rapid cooling and thus leaves unique emission imprints in H$\alpha$ images.
 \end{itemize}
This work provides a good example of possible flux rope creation in the solar sunspot's upper atmosphere, and adds valuable observational constraints to the study of flux rope reconnection.

%% The "ht!" tells LaTeX to put the figure "here" first, at the "top" next
%% and to override the normal way of calculating a float position
\begin{figure}
\centering
\includegraphics[width=0.8\linewidth, keepaspectratio]{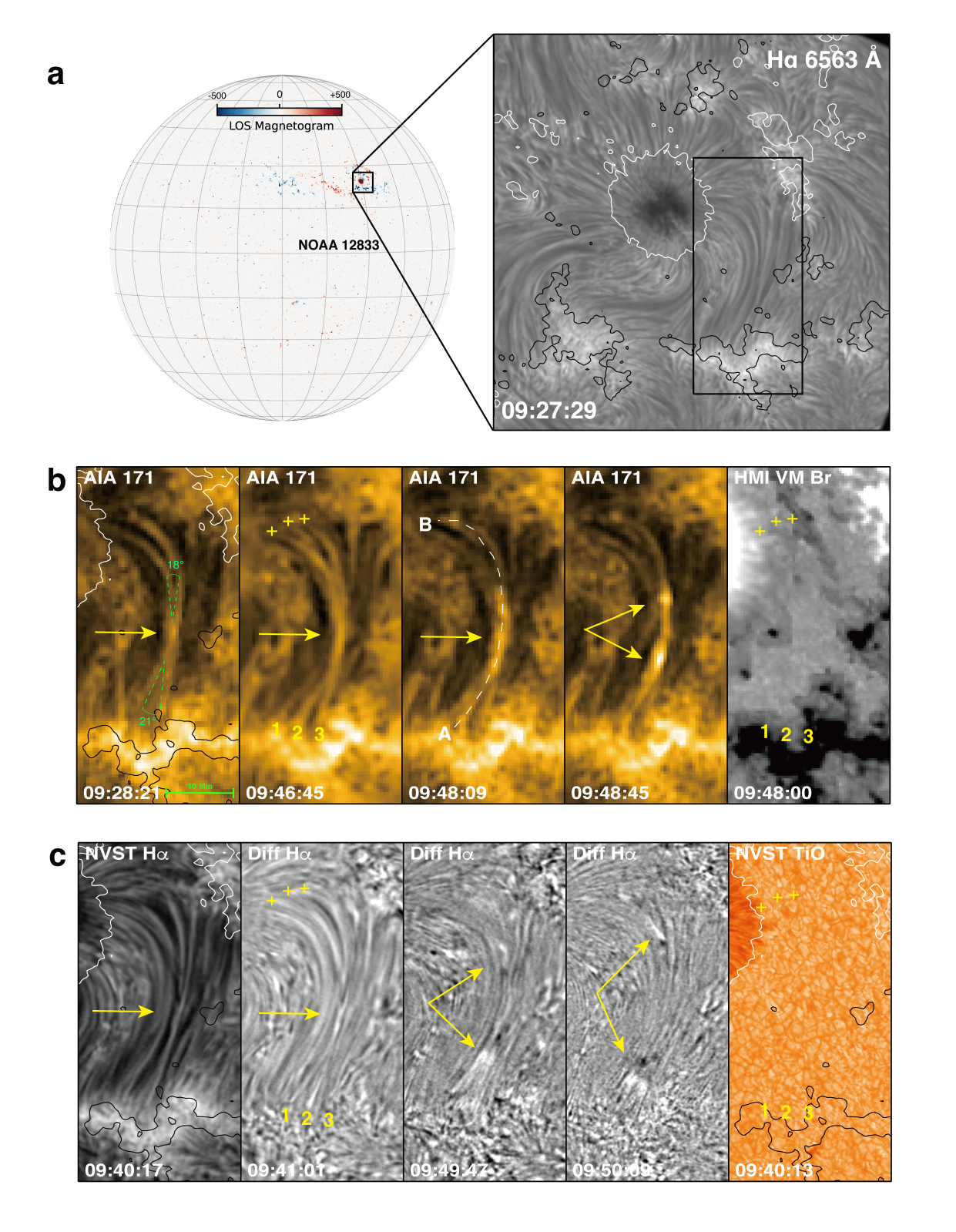}
\caption{Imaging evidence of flux rope reconnection events in AR 12833. (a) The location of AR 12833 in HMI LOS magnetogram and its superpenumbral fibrils in a selected NVST H$\alpha$ image. The black box indicates the field of view (FOV) of panels (b) and (c), which depict the detailed dynamics of  reconnection events and their magnetic field environment. (b) Snapshots in AIA 171 \AA~and HMI radial vector magnetic field with a {saturated value of} $\pm$ 300 G. (c) High-resolution snapshots in NVST H$_{\alpha}$ and TiO images. Three of the H${\alpha}$ images are running difference images, highlighting the braided morphology of chromospheric structures and associated bidirectional outflows.
The white/black contours in panels (a)-(c) outline the photospheric radial vector magnetic field at a strength of ± 300 G, respectively.  In panels (b) and (c),  the footpoints of braided chromospheric/coronal plasma loops are denoted by yellow plus symbols and numbers, respectively. The braided knots of chromospheric/TR loops and their associated bidirectional outflows are marked by yellow arrows. A 12-s animation, combining NVST H$\alpha$ and AIA 171 Å images, showcases the reconnection events during magnetic untangling from 09:22:11 UT to 10:19:45 UT.
\label{fig1}}
\end{figure}

\begin{figure}
\includegraphics[width=\linewidth, keepaspectratio]{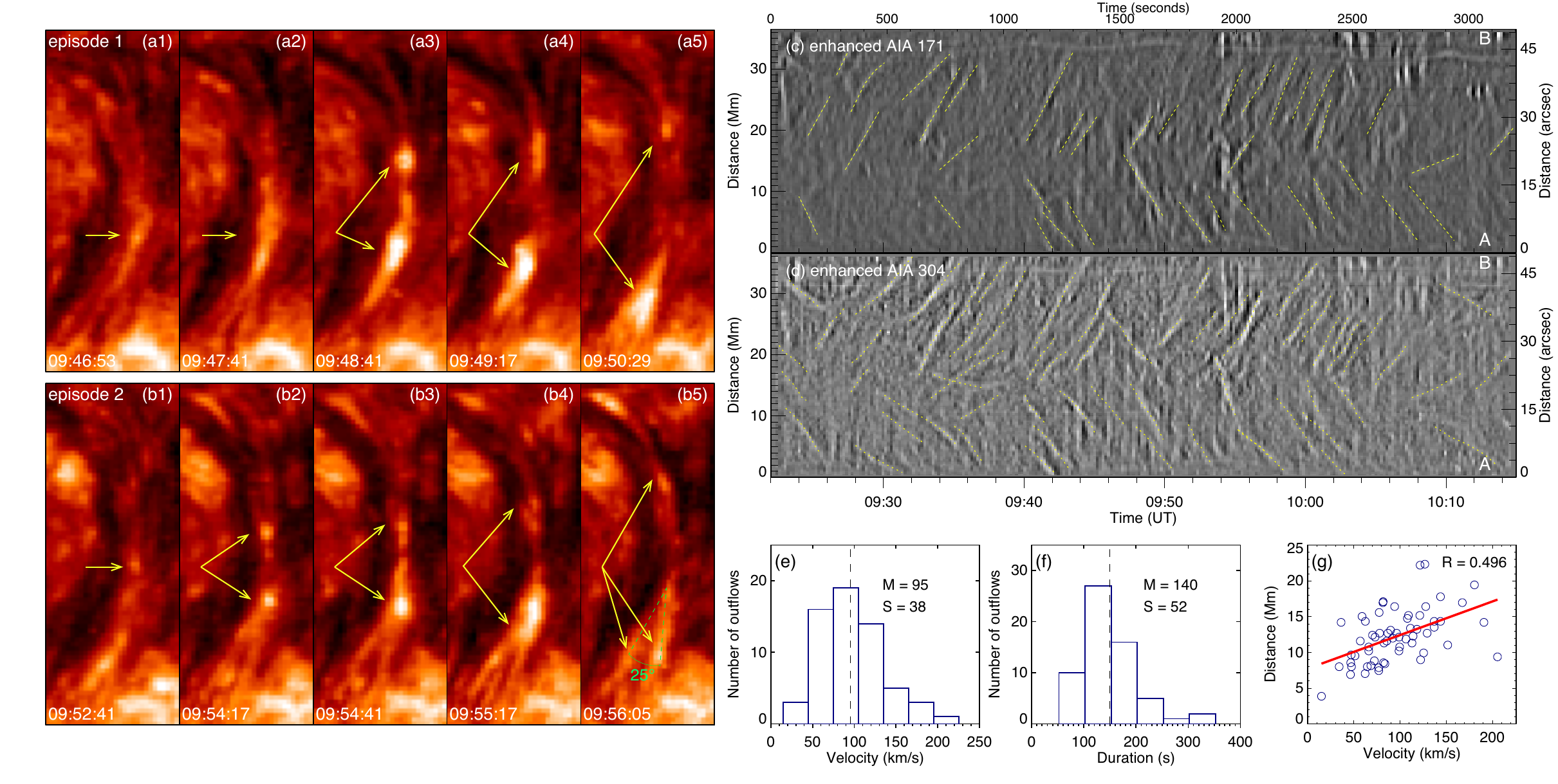}
\caption{Dynamics and statistics of outflows streaming out from reconnection sites. (a1-a4) and (b1-b4): The genesis and dynamics of two episodes of typical outflow blobs observed in AIA 304 passbands. The origin sites of the outflows and moving blobs are marked by yellow arrows. The moving outflow blobs in panels (b3)-(b5) highlight the bifurcated footpoints of braided chromospheric/TR plasma loops. Time-distance plots of {(c)} AIA 171 and {(d)} 304 \AA~images made along {the} slice ``AB" {shown in} the third panel of Figure \ref{fig1}(b).  (e) and (f): The velocity and duration distributions of the identified outflows. (g) The relationship between propagation distance and velocity of identified outflows. An animation of these two episodes is available, covering the full evolution of 304 Å outflow blobs from 09:46:53 UT to 09:56:05 UT.  {The video, with a duration of 4 s, contains both the original and high-pass filtered enhanced 304 images}.
\label{fig2}}
\end{figure}

\begin{figure}
\centering
\includegraphics[width=0.8\linewidth, keepaspectratio]{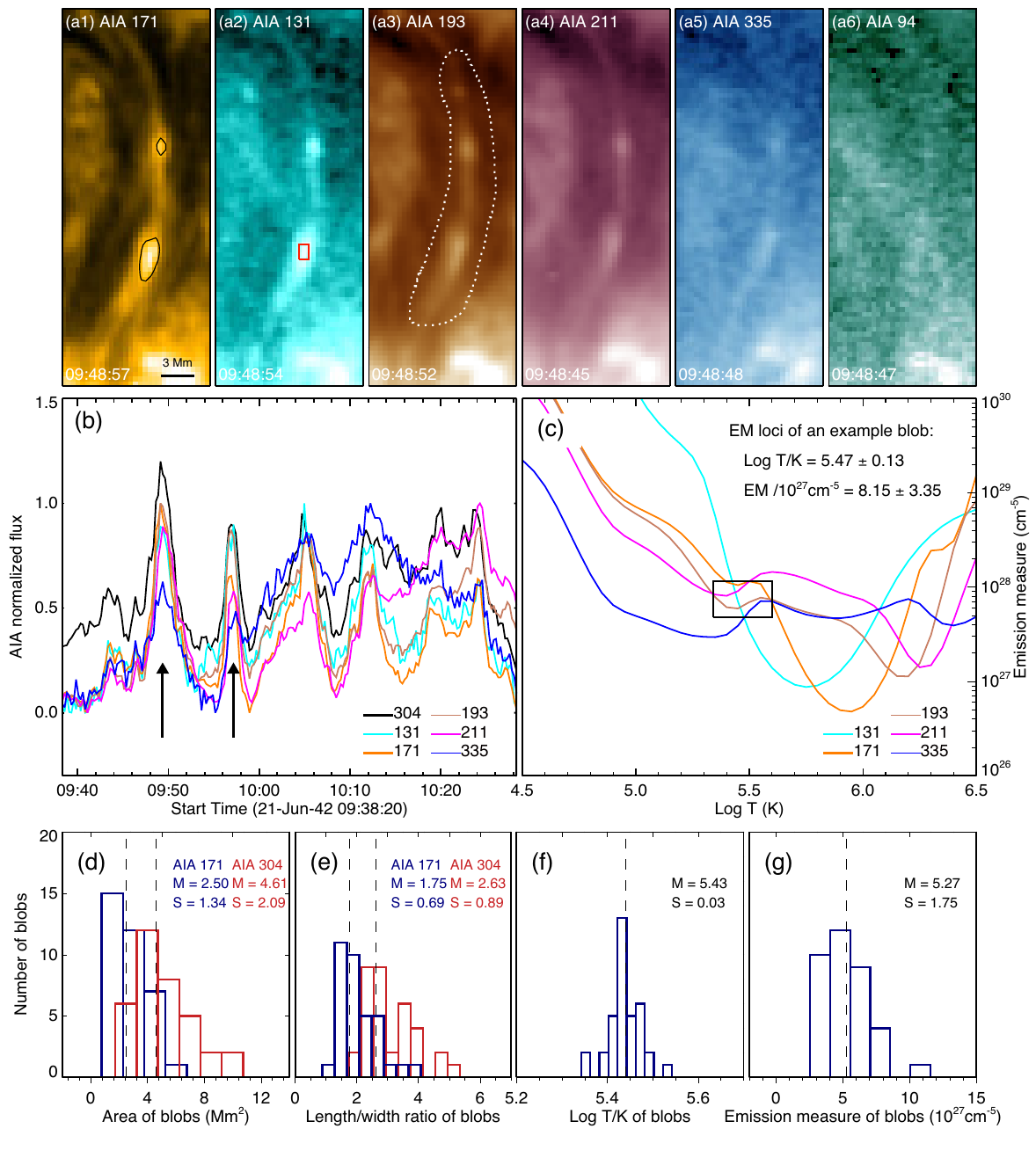}
\caption{{The AIA multi-temperature response of outflowing blobs and their statistical characteristics.} (a1)$-$(a6): Snapshots of typical outflow blobs in AIA 171, 131, 193, 211, 335, 94 \AA~passbands. The black contours outline the identified blob areas in the 171 \AA~image, where the intensity is higher than 3$\sigma$ above the median intensity. (b) The AIA flux variation {at the sites} of the outflowing blobs, outlined by the white dashed line region in panel (a3). (c) The Emission Measure (EM) loci analysis {in} the core of an outflow blob, denoted by the small red rectangle in panel (a2). (d)$-$(g): Statistical analysis of the apparent area, length/width ratio, temperature, and emission measure of the identified outflowing blobs. A 12-second animation, with a 12-second cadence, shows the AIA six-passband response to the outflowing blobs from 09:21:21 UT to 10:19:45 UT, using the same field of view as panels (a1-a6).
\label{fig3}}
\end{figure}

\begin{figure}
\centering
\includegraphics[width=0.8\linewidth, keepaspectratio]{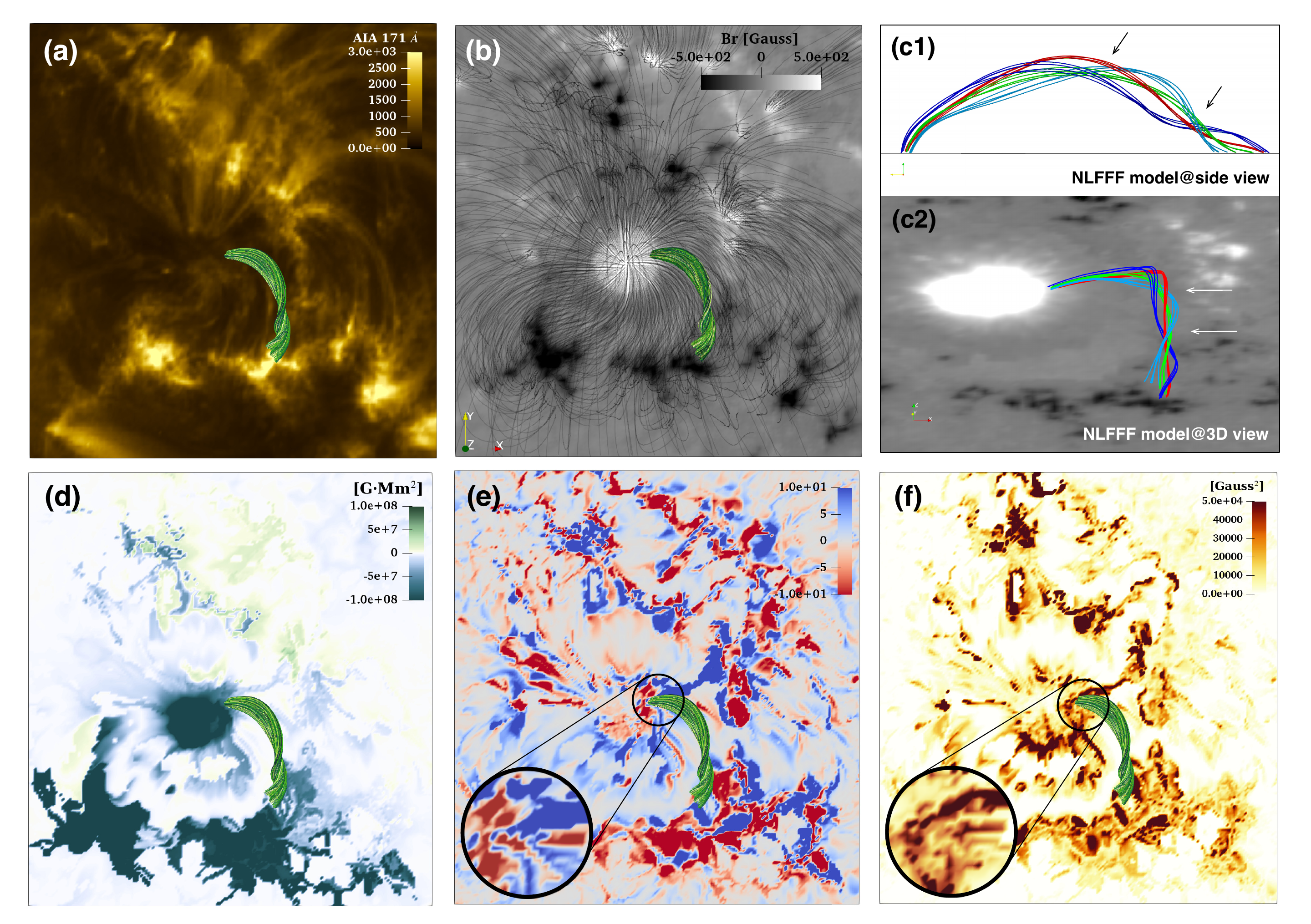}
\caption{{Magnetic field extrapolation and topology analysis.} 
(a) and (d-f): Top views of the NLFFF results with different background images. Black lines represent the background field, and the green-colored lines depict the magnetic field lines of the braided bright threads simulated by the NLFFF model. The background images are the AIA EUV 171 \AA~ taken at 09:46 UT in (a), the radial component of the HMI vector magnetogram taken at 06:12 UT in (b), the field line magnetic helicity in (d), the mean twisting $T(\gamma)$ in  {(d)}, and  the current helicity $H_c(\gamma)$ in (f), respectively. Magnified views of the mean twisting and he current helicity at the northern footpoint region, where the extrapolated magnetic field lines are rooted, are presented in the lower-left insets of panels (e) and (f), respectively. 
(c1) and (c2) present four sets of distinctive colored field lines within the braided structure, shown from side and 3D viewpoints, respectively. Arrows in panels (c1) and (c2) highlight the locations of the braid knots.
\label{fig4}}
\end{figure}

\begin{figure}
\centering
\includegraphics{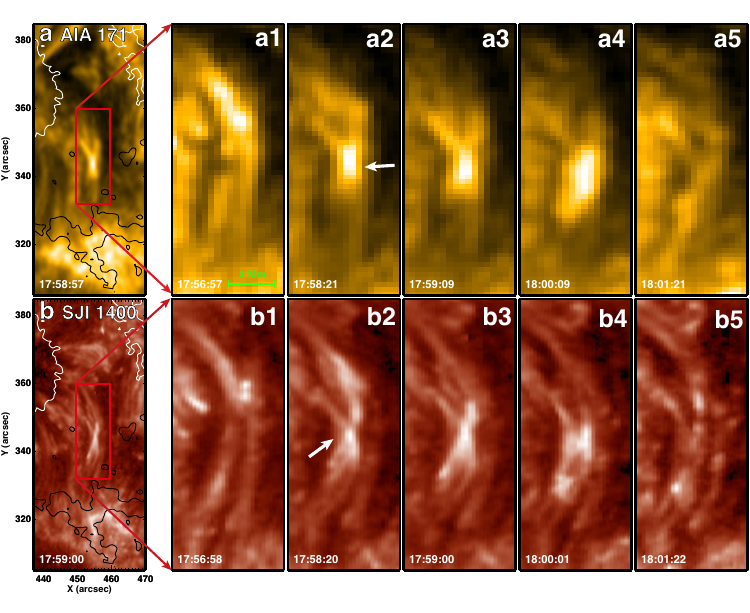}
\caption{{Another episode of flux rope reconnection jointly observed by AIA 171 \AA~ and IRIS SJI 1400 \AA~images.} The panels in the right column are magnified views of the red rectangle in the corresponding panels on the left. The untangling of braided TR plasma loops and associated heating are well-resolved in the IRIS SJI 1400 Å images. The white/black contours in (a) and (b) outline the photospheric radial vector magnetic field at a strength of ± 300 G, respectively.  A 3-second animation shows this reconnection event, covering  {17:56:57 UT to 18:20:57 UT}, using the same field of view as panels (a), (a1), (b), and (b1).
\label{fig5}}
\end{figure}

%% IMPORTANT! The old "\acknowledgment" command has be depreciated. It was
%% not robust enough to handle our new dual anonymous review requirements and
%% thus been replaced with the acknowledgment environment. If you try to 
%% compile with \acknowledgment you will get an error print to the screen
%% and in the compiled pdf.
\begin{acknowledgments}
%H. C. Chen and H. Tian initiated the study. H.C. Chen. processed the primary data and wrote the first manuscript. H. Tian led the discussion and revised the manuscript. E. P. revised the manuscript and improve the scientific implications. C. Xia and X. L. Yan  provided the NVST and participated in the data interpretation.  Z. H. Huang, Z. Y. Hou, and D. Y. Duan contributed to the IRIS data reduction and EM-loci analysis. All authors participated in the discussions and commented the manuscript. 

 {H. C. C. would like to thank Dr. Patrick Antolin for the helpful discussions on the identification of nanojets in our observation.}
H. T. is supported by the NSFC grant 12425301. H. C. C. is supported by the NSFC grant 12573061 and the Yunnan Key Laboratory of Solar Physics and Space Science under the No. YNSPCC202210, as well as the Yunnan Provincial Basic Research Project (202401CF070165).  A. R. Y. and O. E. K. R were supported by UKRI/STFC grant UKRI1216. O.E.K.R and C.P would like to acknowledge funding from UK Science and Technology funding council under grant No. ST/W00108X/1. C. X. was supported by the NSFC grant 12073022, the Strategic Priority Research Program of the Chinese Academy of Sciences (XDB0560000), the Yunnan Key Laboratory of Solar Physics and Space Science (202205AG070009). X. L. Y. is supported by the NSFC grant 12325303. D.Y.D. was supported by NSFC grant 12403065. Z.Y.H. was supported by NSFC grant 12303057. As one of the primary observing facilities of the Fuxian Solar Observatory, NVST is jointly operated and administrated by Yunnan Observatories and Center for Astronomical Mega-Science, Chinese Academy of Science. IRIS is a NASA Small Explorer mission developed and operated by LMSAL with mission operations executed at NASA Ames Research center and major contributions to downlink communications funded by ESA and the Norwegian Space Center. SDO is a mission for NASA’s Living With a Star (LWS) program. 
\end{acknowledgments}

%% To help institutions obtain information on the effectiveness of their 
%% telescopes the AAS Journals has created a group of keywords for telescope 
%% facilities.
%
%% Following the acknowledgments section, use the following syntax and the
%% \facility{} or \facilities{} macros to list the keywords of facilities used 
%% in the research for the paper.  Each keyword is check against the master 
%% list during copy editing.  Individual instruments can be provided in 
%% parentheses, after the keyword, but they are not verified.

\vspace{5mm}

%% Similar to \facility{}, there is the optional \software command to allow 
%% authors a place to specify which programs were used during the creation of 
%% the manuscript. Authors should list each code and include either a
%% citation or url to the code inside ()s when available.

%% Appendix material should be preceded with a single \appendix command.
%% There should be a \section command for each appendix. Mark appendix
%% subsections with the same markup you use in the main body of the paper.

%% Each Appendix (indicated with \section) will be lettered A, B, C, etc.
%% The equation counter will reset when it encounters the \appendix
%% command and will number appendix equations (A1), (A2), etc. The
%% Figure and Table counter will not reset.

\appendix
\renewcommand{\thefigure}{A\arabic{figure}} % Prefix figures with "A" for the appendix
\setcounter{figure}{0} % Reset the figure counter for the appendix
\section{ {Weak Nanojets detected in flux rope reconnection}}

 {Nanojets are collimated, small-scale plasma ejecta that have been recently reported in high-resolution solar observations \citep[e.g.,][]{2021NatAs...5...54A,2022ApJ...938..122P,2022ApJ...934..190S,2025ApJ...985L..12G}. In the scenario of flux rope reconnection, they are expected to be triggered perpendicular to the reconnecting field line.
In this study, we indeed detected AIA signatures of ongoing nanojets as repetitive reconnection events occurred inside the twisted and braided flux rope (see Figures \ref{figA0} and its animation). In contrast to the previously investigated field-aligned bidirectional outflowing blobs, the dynamics and physical properties of these weak nanojets are challenging to study without higher-resolution imaging. This study, therefore, only reports these weak events as evidence of ongoing flux rope reconnection.}

 {As shown in Figures \ref{figA0}(c1-c3) and (d1-d2), Nanojet 1 and Nanojet 2 emanated from the emission-enhanced region where the bright threads crossed with each other. 
These nanojets were found to be very weak, and could only be discerned in 304 Å images during several episodes of such repetitive flux rope reconnection events. They were short-lived, rapid features with a clear velocity component perpendicular to the reconnecting bright threads. The nanojets were ejected along two distinct paths, marked as "slit 1" and "slit 2," with projective velocities  of 62 km s$^{-1}$ and 40 km s$^{-1}$, respectively. An accompanying animation clearly demonstrates their occurrence, where yellow arrows mark the ongoing nanojets at each time frame. The characteristics of the nanojets in our event are consistent with previous works \citep[e.g.,][]{2021NatAs...5...54A,2022ApJ...938..122P,2022ApJ...934..190S,2025ApJ...985L..12G,2025arXiv250904741T}.}

\begin{figure}[ht!]
\centering
\includegraphics[width=0.8\linewidth, keepaspectratio]{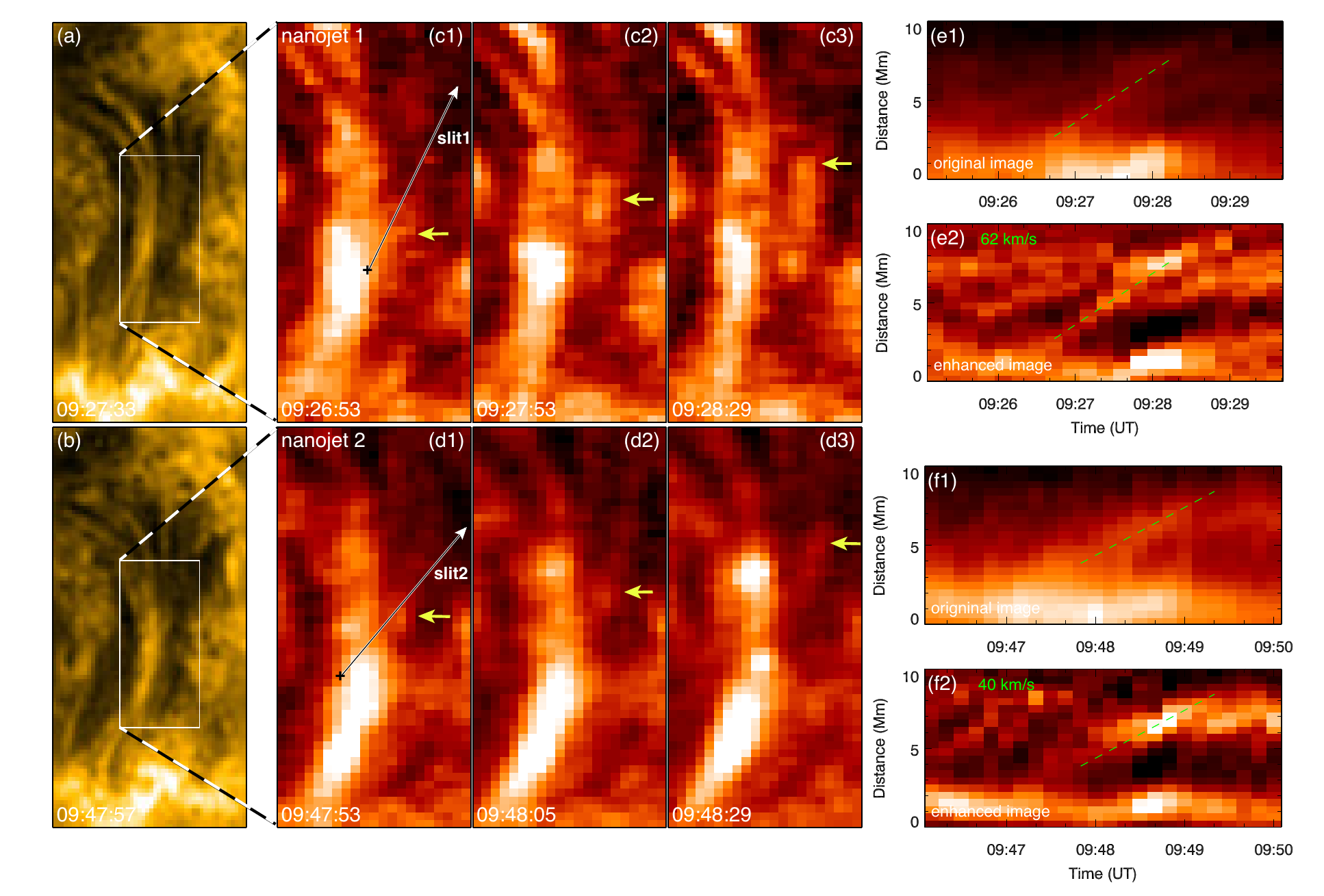}
\caption{ {304 \AA~nanojets detected in the flux rope reconnection.  Panels (a) and (b) show the braided and twisted bright threads observed in the 171 \AA~passband. These threads outline the magnetic configuration of the flux rope and contain the reconnection region (outlined by the white rectangle) where the nanojets were detected. Panels (c1-c3) and (d1-d3) show two examples of these nanojets, generated from the middle crossing site of the bright threads during the time periods of 09:27–09:30 UT and 09:47–09:49 UT, respectively. The moving nanojets are marked with yellow arrows. Panels (e1-e2) and (f1-f2) show the original and high-pass filtered enhanced AIA 304 Å time-distance diagrams along slit 1 and slit 2, respectively. We measured the velocities of the nanojets by linear fitting on these diagrams.}}
\label{figA0}
\end{figure}

\section{photospheric tangential velocity fields at flux rope footpoints}

Figure A2 presents a 3D view of NLFFF magnetic extrapolation of the source region, and the photospheric tangential velocity fields at flux rope footpoints.

\begin{figure}[ht!]
\centering
\includegraphics[width=0.8\linewidth, keepaspectratio]{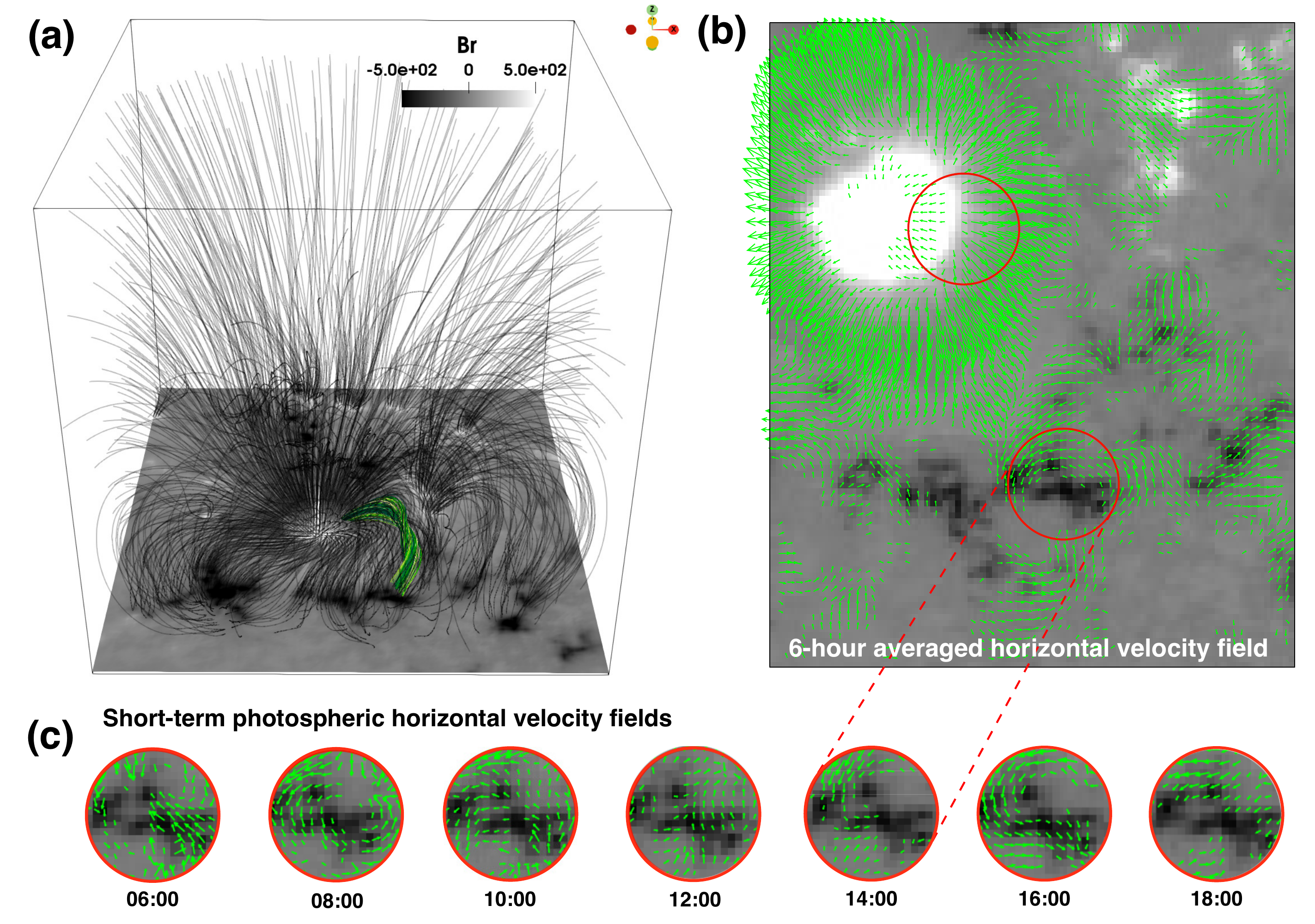}
\caption{(a) 3D view of the NLFFF results.  The background images display the radial component of the vector magnetogram obtained by HMI at 06:12 UT, saturated at ± 500 G. (b) Long-term photospheric horizontal flow fields, as depicted by green arrows, averaged over the time period from 06:00 UT to 12:00 UT, with a maximum and median horizontal velocities of 1.30 km s$^{-1}$ and 0.20 km s$^{-1}$, respectively.  Red circles indicate the footpoint regions of the bright threads. (c) Short-term photospheric horizontal flow fields, derived over a 12-minute period, at the south footpoint region of the bright threads.}
\label{figA1}
\end{figure}

Using HMI vector magnetograms, the long-term and short-term photospheric tangential plasma velocity fields were derived for the footpoint regions of the flux rope reconnection event, with the differential affine velocity estimator for vector magnetograms \citep[DAVE4VM,][]{2008ApJ...683.1134S}. At the southern footpoint region of the flux rope, the 12-minute cadenced results reveal more random and complex flow patterns that varied over shorter time periods.  As shown in Figure \ref{figA1} (b) and (c), the two footpoint regions of the extrapolated flux rope (marked by two red circles) are characterized by long-term, near-paralleled horizontal flows, namely northwest sunspot moat flows and southeast supergranular flows, which may shear the corresponding footpoints of the coronal field lines. At the southern footpoint region of the flux rope, the 6-hour averaged tangential plasma velocity fields indicate a persistent anti-clockwise vortex flow from 06:00 UT to 12:00 UT. Conversely, the 12-minute cadenced results reveal more random and complex flow patterns that vary over shorter time periods. This complexity, rather than a simple vortex, near the flux rope's southern footpoint, could effectively create a braided flux rope.  Future research will need to delve deeper into the relationship between the slow photospheric motions and the magnetic field tangling, particularly regarding the production of braiding. This will be promising with the availability of higher-resolution observations from upcoming new-generation solar telescopes, including large-aperture instruments like the Daniel K. Inouye Solar Telescope in the U.S., the Chinese 2.5-meter Wide-Field High-Resolution Solar Telescope, and the proposed 8-meter Chinese Giant Solar Telescope.

%% For this sample we use BibTeX plus aasjournals.bst to generate the
%% the bibliography. The sample631.bib file was populated from ADS. To
%% get the citations to show in the compiled file do the following:
%%
%% pdflatex sample631.tex
%% bibtext sample631
%% pdflatex sample631.tex
%% pdflatex sample631.tex

\bibliography{sample.bib}
\bibliographystyle{aasjournal}

%% This command is needed to show the entire author+affiliation list when
%% the collaboration and author truncation commands are used.  It has to
%% go at the end of the manuscript.
%\allauthors

%% Include this line if you are using the \added, \replaced, \deleted
%% commands to see a summary list of all changes at the end of the article.
%\listofchanges

\end{document}